# The postulates of Quantum Mechanics*

V. Dorobantu,

Physics Department, Politehnica University, Timisoara, Romania

*Abstract*
   As a starting point in understanding Quantum Mechanics, the postulates of Quantum Mechanics are presented, and few of the main eigenvalue problems, as well.

**Introduction**.
   Quantum Mechanics is an axiomatic theory because it is well-grounded on few principles (from the Latin *principium*), or axioms (from the Greek, *axios*), or postulates (from the Latin *postulatum*), all of these words meaning the same thing: a truth which doesn't need any further proof, because it is obvious by itself.

There is not a consensus of how many axioms one needs to describe the machinery of Quantum Mechanics, but I think that five is an appropriate number. The first four postulates, as we shall see, make up the mathematical background of Quantum Mechanics, and the fifth supplies the connection between the mathematics introduced by the first four and the results of a measurement process.

As a general rule, the statement of each postulate will be followed by comments, so that the significance of words within the postulates will be explained at the right time, namely, when they are introduced.

## 1. The first postulate of Quantum Mechanics

***To every state of a physical system there is a function $\Psi$ ascribed to and defining the state.***

First of all: "*there is a function $\Psi$ ascribed to…*" doesn't mean that there is a one-to-one correspondence function – state. As we will see, a state may have more than one function if certain conditions are fulfilled.

"*To define means to limit*" according to the English writer, Oscar Wilde, so our definitions will be closer to the usual vocabulary in order to make the book more accessible, but it doesn't mean that any accuracy is lost.

A **physical system** is (from the quantum point of view) a free particle, a particle moving in some potential, a hydrogen atom, a hydrogen molecule, or an atom (or molecule) of whatever kind, or many particles of the same kind or different. As we can see, by a physical system we can understand a finite region of space having certain characteristics which make that region different from others.

If there are certain physical quantities, or parameters, which at least in principle can be measured, and they remain constant for a finite time interval, then we can speak about the *state* of the physical system. It should be noted that there is a difference between a state from the classical point of view and the quantum point of view:

a) let us say we have a thermodynamic state of an ideal gas, meaning that *we know* (for sure) the pressure, the temperature and the volume. It is understood that we have to do with a maximum specification when we speak of state;

b) things are different from the quantum point of view when we speak of state, because the verb *to know* does not have the meaning of 'we know for sure,' but, as we shall see, *we know with certain probabilities*.

To make a connection, we will speak about a state with maximum specification only when that state is a *pure* state, and less then maximum specification when we deal with a *mixed* (or an "*impure*") state. What do we mean by *pure* and *mixed* states, we will find out a little bit later.

`Ψ(x,y,z,t)`, depending of the usual space coordinates (x, y, z) and time t, **is called a wavefunction.** When we speak about the *state of a physical system*, then Ψ *depends of space coordinates* only, and when we speak about *evolution, the time dependence of* Ψ must be taken into account. Generally, the wave-function is a complex function, which means that we have to do with its complex conjugate $\Psi^*$, as well, obtained by changing the imaginary number i into -i.

**The significance of the wave-function**

Max Born has the merit of clarifying the meaning of the ***wave-function*** as being ***the probability amplitude.*** Let us consider the simplest atom consisting of one proton and one electron – the hydrogen atom. Taking the proton as the centre of a reference system, the hydrogen's electron must be somewhere around the proton at, say the distance r. The wave.-function will be $\Psi(x,y,z) = \Psi(r,\theta,\varphi)$, if we work in spherical coordinates. The probability to find the electron somewhere at a distance r from the proton is:

$$\int_0^r |\Psi(r,\theta,\varphi)|^2 dV = \int_0^r |\Psi(r)|^2 r^2 \sin\theta d\theta d\varphi dr =$$

$$4\pi \int_0^r |\Psi(r)|^2 r^2 dr \qquad (1)$$

and the *probability density* is:

$$\rho_\Psi(r) = 4\pi r^2 \Psi^*(r) \Psi(r) = 4\pi r^2 |\Psi(r)|^2 \qquad (2)$$

If we take the integral of expression (1), from zero to infinity, we are sure that the electron will be in this sphere, and, as a consequence the probability will be 1.

Considering $x = \{x_1, x_2, x_3, \ldots \ldots x_n\}$, the generalized coordinates [1],[28], or Lagrangean parameters, the probability to find the particle in the entire space must be 1, so:

$$\int_{\text{all space}} |\Psi(x)|^2 dV = 1 \qquad (3)$$

where $dV = dx_1 dx_2 dx_3 \ldots\ldots dx_n$ is the elementary volume in the configuration space (the space of generalized coordinates).

An expression like (3) defines a special class of functions of unit norm, meaning that all $\Psi$'s have a modulus of one.

*How does one work with the probability amplitude $\Psi$ ?*

Let us see Fig. 1, with electrons leaving the source EG and reaching the detector D. The probability amplitude for this process is $\Psi_{(EG)D}$.

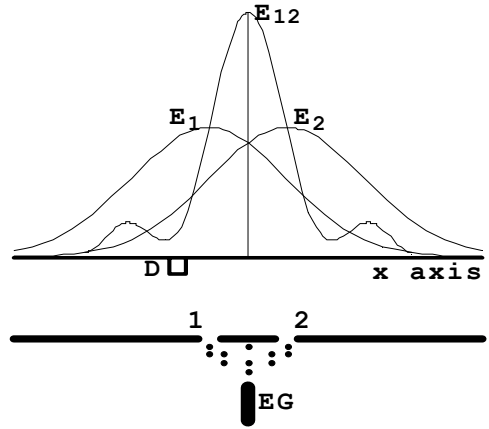

Fig.1

The particle can take the path through slit 1, or through slit 2 and the probability amplitude for these processes is $\Psi_{(EG)1D} = \Psi_{(EG)1} \Psi_{1D}$, respectively $\Psi_{(EG)2D} = \Psi_{(EG)2} \Psi_{2D}$ with:

$$\Psi_{(EG)D} = \Psi_{(EG)1} \Psi_{1D} + \Psi_{(EG)2} \Psi_{2D} \tag{4}$$

and the density probability:

$$\left|\Psi_{(EG)D}\right|^2 = \left|\Psi_{(EG)1}\Psi_{1D} + \Psi_{(EG)2}\Psi_{2D}\right|^2 \tag{5}$$

Another way of writing the probability amplitude is to use Dirac's *bra* $\langle\ |$ and *ket* $|\ \rangle$ vectors.

First of all, **why vectors**? Because there is *another name* for the wave-function, or probability amplitude, namely *state vector*. Let us have a vector $\vec{r}$ of modulus 1, and $\vec{i}, \vec{j}, \vec{k}$ the unit vectors of Ox, Oy and Oz axes. Let also $\theta_x, \theta_y, \theta_z$ be the angles made by $\vec{r}$ with Ox, Oy and Oz, see Fig. 2.

$$\vec{r} = x\vec{i} + y\vec{j} + z\vec{k}, \quad x = \cos\theta_x, y = \cos\theta_y, z = \cos\theta_z \tag{6}$$

with $r^2 = x^2 + y^2 + z^2$ and r having modulus 1 we will get:

$$\cos^2\theta_x + \cos^2\theta_y + \cos^2\theta_z = 1 \tag{7}$$

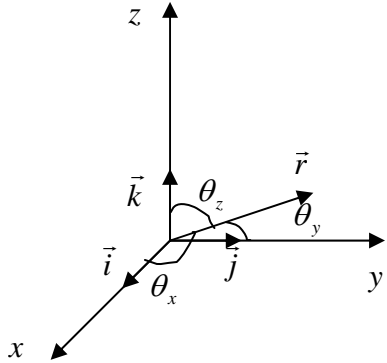

Fig. 2

A straightforward generalization of expression (6) is to consider an n – dimensional space spanned over the base unit
vectors $\Psi_1, \Psi_2, \Psi_3, \ldots \ldots \Psi_n$. Then each other vector of modulus 1 can be written as:

$$\Psi = \sum_{i=1}^{n} c_i \Psi_i \qquad (8)$$

with $c_k$ being complex coefficients satisfying a condition similar to (7)

$$\sum_{k=1}^{n} |c_k|^2 = 1 \qquad (9)$$

Actually, the expression (8) is the mathematical formulation of a general physical principle, namely, ***the superposition principle*** stating that *if $\Psi_1, \Psi_2, \Psi_3, \ldots \ldots \Psi_n$ represent physically realizable processes, then any linear combination of the form* (8) *is also a physical realizable process.*
A system of functions allowing the expression (8) and the closure relation (9) is called complete and closed.

*Dirac's writing*.
Splitting the word *bracket* in two: *bra* and *ket*, Dirac invented a special form of writing, generally used today in Quantum Theories. Here is Dirac's writing: a physical process starting from an initial state and reaching a final state is described by the probability amplitude

$$\langle \text{final} | \text{initial} \rangle \qquad (10)$$

The vector $\langle \text{final} |$ is the *bra* vector and $| \text{initial} \rangle$ is the *ket* vector, observing that the bra vector $\langle \text{final} |$ is the complex conjugate of the ket vector $| \text{final} \rangle$. What we have in the right part of the vertical line is always the initial state, and what we have in left part is the final state. For instance, $\Psi_{(EG)D}$, the probability amplitude describing the process of a particle leaving the source EG and reaching the detector D is now written as:
$$\Psi_{(EG)D} = \langle D|EG\rangle \tag{11}$$
and the equivalent of formula (4) is :
$$\langle D|EG\rangle = \langle D|1\rangle\langle 1|EG\rangle + \langle D|2\rangle\langle 2|EG\rangle \tag{4'}$$
reading the expressions from the right to the left as Feynman said [2]: the particle leaves the source EG, passes through orifice 1 (or 2) and reaches the detector D. $\langle 1|EG\rangle$ is the probability amplitude for the process of leaving the source and reaching orifice 1, and $\langle D|1\rangle$ the probability amplitude for the process of leaving the orifice 1 and reaching the detector D, their product giving the probability amplitude for the process EG – 1- D.

*Conclusion*
1. We saw that the *wave-function*, the *probability amplitude* and the *state vector*, represent the *same* thing.
2. We were introduced to *functions such as*, $\Psi$ *, which have the following properties*:
a) the probability amplitudes are *uniform* functions, namely in every point of the physical space have a definite value;
b) the wave – functions are *bounded* in entire space;
c) the wave- functions are *continous* and their *partial derivatives* are also *continous*;
d) the functions are integrable in squared modulus, namely
$\int |\Psi|^2 dV$ is *convergent*;

*Functions with properties a) – d) make up Hilbert's complex linear vector space of wave- functions.*
3. To every pair of the probability amplitudes $|\Phi\rangle$ and $|\Psi\rangle$ from the Hilbert space is associated a number called *inner product* $\langle\Phi|\Psi\rangle$ defined as:
$$\langle\Phi|\Psi\rangle = \int \Phi^* \Psi \, dV \tag{12}$$
The inner product:
 (i) is non-comutative or skew-symetric:
$$\langle\Phi|\Psi\rangle \neq \langle\Psi|\Phi\rangle$$
$$\langle\Phi|\Psi\rangle = \langle\Psi|\Phi\rangle^* \tag{13}$$
 (ii) has the property of linearity:
$$\langle\Phi|(a|\Psi_1\rangle + b|\Psi_2\rangle)\rangle = a\langle\Phi|\Psi_1\rangle + b\langle\Phi|\Psi_2\rangle \tag{14}$$
  with a and b two constants and $|\Psi_1\rangle$
 (iii) has the property of positivity:
$$\langle\Psi|\Psi\rangle \geq 0 \tag{15}$$
$$\langle\Psi|\Psi\rangle = 0, \text{ only if } |\Psi\rangle = 0$$

4. *Orthogonality.* If the inner product of the two wave-functions is null, then those two wave-functions are *orthogonal*. Here, the word *orthogonal* is not too far from the usual meaning, namely: perpendicularity. In Fig. 2 we have examples of orthogonality: the unit vectors $\vec{i}, \vec{j}, \vec{k}$ are each other perpendicular. To see how it works, let's write the complex conjugate of the expression (8):

$$\langle \Psi | = \sum_k c_k^* \langle \Psi_k | \tag{16}$$

Taking the inner product with $|\Psi\rangle$, we have:

$$\langle \Psi | \Psi \rangle = \sum_k c_k^* \langle \Psi_k | \Psi \rangle = \sum_k c_k^* c_k \tag{17}$$

because wave-functions $|\Psi\rangle$ have the norm 1, see (3), and the complex coefficients $c_k$ satisfy the closure relation (9). By comparing coefficients in (17) we get:

$$c_k = \langle \Psi_k | \Psi \rangle \tag{18}$$

According to expression (18), the *coefficients of the* $|\Psi\rangle$ *expansion* (8) *are uniquely determined*. Introducing the $|\Psi\rangle$ development in (18) we have:

$$c_k = \left\langle \Psi_k \left| \sum_j c_j \right| \Psi_j \right\rangle = \sum_j c_j \langle \Psi_k | \Psi_j \rangle \tag{19}$$

For the expression (19) to be satisfied,

$$\langle \Psi_k | \Psi_j \rangle = \delta_{kj}, \quad \delta_{kj} = \begin{cases} 1, k = j \\ 0, k \neq j \end{cases} \tag{20}$$

where $\delta_{kj}$ is the Kronecker symbol.

The above expression, (20) represents the general relation of *orthonormality* : the inner product is 1 for normalized wave-functions, and zero for orthogonal wave-functions. Any collection of n ($|\Psi_1\rangle$, $|\Psi_2\rangle$,....$|\Psi_n\rangle$) mutually orthogonal vectors of unit length in an n-dimensional space vector, satisfying condition (20), makes up the *orthonormal* basis for that space.

*Note*. The first postulate doesn't tell us how to find the wave-function, and, as matter of fact, it is not meant to tell us that. This postulate only says that the wave-function offers the maximum knowledge. Regarding the probability amplitude, or wave-fuction, or state vector, and what are they good for, we shall learn more after the third postulate.

**2. The second postulate of Quantum Mechanics**

*If $\mathcal{H}_1$ is the Hilbert space associated with the physical system S₁, and $\mathcal{H}_2$ is the Hilbert space corresponding to the other physical system S₂, then the composite system S₁ + S₂ will be associated with the tensor product of the two Hilbert vector spaces $\mathcal{H}_1 \otimes \mathcal{H}_2$.*

As a summary, the second postulate of Quantum Mechanics defines the word **and,** as Roger Penrose [3] suggests. Let us consider the non-null vectors $\Psi_1, \Psi_2 \in \mathcal{H}_1$ and $\Phi_1, \Phi_2 \in \mathcal{H}_2$. The state vector $\Psi_1 \otimes \Phi_1$ is going to tell us the following: the system S₁ is in state $\Psi_1$ **and,** at the same time, the system S₂ is in state $\Phi_1$. This is similar for $\Psi_2 \otimes \Phi_2$. We have to note that the concept of tensor product, allowing the presence of both S$_1$ **and** S$_2$ at the same time, it is completely different from the linear superposition where if, say, $|\alpha\rangle$ and $|\beta\rangle$ are two possible states for *one* particle, then a linear combination of the form (8) will also a possible state for the same particle, but *not* for *two* paricles. Having in mind the second postulate, the state for *two* particles is described by the tensor product. Now, the vectors $\Psi_1, \Psi_2 \in \mathcal{H}_1$ are of unit length, and they are *orthogonal*. This also holds for $\Phi_1, \Phi_2$. What about the new unit vector $\Psi$?

$$\Psi = \frac{1}{\sqrt{2}}\left(\Psi_1 \otimes \Phi_1 + \Psi_2 \otimes \Phi_2\right) \qquad (21)$$

$\Psi$ is a vector in $\mathcal{H}_1 \otimes \mathcal{H}_2$ describing a state of the composite system S$_1$ + S$_2$. *When the composite system* S$_1$ + S$_2$ *is in the state* $\Psi$, *can we say in what state* S$_1$ *is in, and what state* S$_2$ *is in?* Definitely not! *Neither* S$_1$, *nor* S$_2$ *are in a definite state*.
*The state* $\Psi$ *given by* (21) *is an **entangled** state* [4].
This second postulate outlines another strange possibility: *quantum non-locality.* If both systems S$_1$ and S$_2$ are well separated in space, the entangled state $\Psi$ of S$_1$ + S$_2$ is a manifestation of non-locality, but a non-locality in a correlated manner, because if we measure a physical quantity for the first particle, the same physical quantity for the second seems to be already fixed by the measurement done on the first, as we shall see later.
*The entangled states and the quantum non-locality are very important for quantum computation.*

### 3. The third postulate of Quantum Mechanics

*To every observable of a physical system is associated a self-adjoint (or Hermitian) operator allowing a complete set of eigenfunctions.*

An **observable** is any physical quantity which can be measured by an experimental procedure, such as: *position, momentum, angular momentum, energy,* etc.
An **operator** is an instruction showing us how to obtain a function g(x) if we know another function f(x), or *the operator maps a function* f(x) *into another function* g(x). From now on the operators will be written with capital bold letters. Let **O** be an operator and f(x) a certain function of coordinate x. Applying the operator **O** to the function f(x) we will get the function g(x):

$$\mathbf{O} f(x) = g(x) \qquad (22)$$

Being in the frame of Quantum Mechanics, *our functions are the wave-functions,* $\Psi$ *'s, so that the operators will map a function from Hilbert space into another function from Hilbert space.*
The *physical equivalent of an operator* is some device changing something in a particle's state. For instance: an electron moves along a straight line. If the electron enters the electric field of a parallel-plate capacitor, its path will be curved, so the electron is in another state. In this case, the *parallel-plate capacitor* is the *physical operator*.

Let $\mathcal{B}$ be an observable, **B** the corresponding operator and $\Phi, \Psi$ two vector states from Hilbert space; then according to (22) we can write:
$$\mathbf{B}\Psi = \Phi \tag{23}$$
If $\Phi$ can be written as $b\Psi$, with b a number, then the equation:
$$\mathbf{B}\Psi = b\Psi \tag{24}$$
is the *eigenvalue equation* of the operator **B**, $\Psi$ being *eigenfunctions* and b the *eigenvalues* of **B**.
Let b and f be constants and $\Phi, \Psi$ two wave-functions.
**B** is a *linear operator* if:
$$\mathbf{B}(b\Psi + f\Phi) = b\mathbf{B}\Psi + f\mathbf{B}\Psi \tag{25}$$
Since we deal with linear operators only, as a general rule, we shall omit the word *linear*.
*The theory of linear operators is the mathematical apparatus of Quantum Mechanics.*

Examples

Log is an operator, $\sqrt{\phantom{x}}$ is also an operator, but both, **Log** and $\sqrt{\phantom{x}}$ *are not linear* operators because $\text{Log}(b\Psi + f\Phi) \neq b\text{Log}\Psi + f\text{Log}\Phi$ and, also, $\sqrt{b\Psi + f\Phi} \neq b\sqrt{\Psi} + f\sqrt{\Phi}$, because they do not satisfy rule (25)
**Derivative** and **integral** are *linear* operators because they satisfy rule (25) being distributive against addition.

*Eigenvalues spectra*
The general problem of Quantum Mechanics is to solve equations of the type (24), or, in other words, to get the solutions for eigenfunctions $\Psi$, and eigenvalues b.
Regarding the eigenvalues, the *ensemble* of the b values we have got as *solutions of the linear operator equation* makes up the *eigenvalues spectrum*. If b can take only *certain* values, we say that the eigenvalues *spectrum is discrete*. If b can take *any* value, then the *spectrum is continuous*. For example, the position of a particle can take any value along the axis of real numbers, so its eigenvalues spectrum is a continuous one.
The hydrogen's electron can be only on certain levels, so the electron's energy can take certain values only, its eigenvalues spectrum being discrete.
If the *eigenvalue spectrum* is a *discrete* one, then the *norm* is of the type (3) or (20) and the expansion using the basis vectors is of the form (8) with closure relation (9), but if the spectrum is a continuous one, such kind of expressions are, as we shall see, of a different type.
For some details of continuous spectra, see, e.g. [5,6].

*A few necessary definitions*
a) If we have two operators **A** and **B**, then their *sum* **A** + **B** is also an operator:
$$(\mathbf{A} + \mathbf{B})\Psi = \mathbf{A}\Psi + \mathbf{B}\Psi \tag{26}$$
b) If we have two operators **A** and **B**, then their *product* **A B** is also an operator:
$$(\mathbf{A}\,\mathbf{B})\Psi = \mathbf{A}(\mathbf{B}\Psi) \tag{27}$$
c) If we have two operators **A** and **B**, then:
$$[\mathbf{A},\mathbf{B}] = \mathbf{A}\mathbf{B} - \mathbf{B}\mathbf{A} \tag{28}$$
is also an operator and if

$$[\mathbf{A},\mathbf{B}] = \mathbf{A}\mathbf{B} - \mathbf{B}\mathbf{A} = 0 \tag{29}$$

then we say that the two operators **A** and **B** *commute* and **[A,B]** is called the commutator of **A** and **B**

d) If **A** is an operator and if there is an operator **A**⁻¹, such that

$$\mathbf{A}\mathbf{A}^{-1} = \mathbf{A}^{-1}\mathbf{A} = \mathbf{I} \tag{30}$$

then **A**⁻¹ is the *inverse* operator, with **I** the *identity* operator.

*Self-adjoint operator*

While studying a physical process, we are interested in values: values of energy, values of angular momentum, etc., namely in real numbers. The corresponding *operators dealing* with *real numbers* (as eigenvalues) are *self-adjoint* or *Hermitian operators.* Let's see how it works!

Let **A** be an operator and $\Psi$, $\Phi$ two state vectors. If the following equality takes place:

$$\langle \Psi | (\mathbf{A} | \Psi \rangle) \rangle = \langle (\langle \mathbf{A}^+ \Psi |) | \Phi \rangle \tag{31}$$

then **A**⁺ is called the *adjoint* operator of **A.**

If
$$\mathbf{A} = \mathbf{A}^+ \tag{32}$$

then **A** is a *self-adjoint*, or **Hermitian** operator.

*The eigenvalues of self-adjoint operators are real numbers.*

Let **A** be a Hermitian operator and $\Psi$ one of its eigenfunctions, then according to (32) we can write:

$$\langle \Psi | \mathbf{A}\Psi \rangle = \langle \mathbf{A}\Psi | \Psi \rangle \tag{33}$$

With $\mathbf{A}\Psi = a\Psi$, equation (51) becomes:

$$\langle \Psi | a\Psi \rangle = \langle a\Psi | \Psi \rangle \Rightarrow a\langle \Psi | \Psi \rangle = a^* \langle \Psi | \Psi \rangle \Rightarrow$$
$$(a - a^*)\langle \Psi | \Psi \rangle = 0, \langle \Psi | \Psi \rangle \neq 0 \Rightarrow a = a^* \Rightarrow a \text{ is real} \tag{34}$$

*Observation*. I saw the following sentence in a university book [7, p38]: "In quantum mechanics, an *observable* is a *self-adjoint operator*". It seems to me the observation is somewhat shallow. An *observable* is a physical quantity, and an *operator* is a mathematical "device" acting on a vector space, as the third postulate clarifies.

*The momentum and position eigenvalue problems*

Momentum and position are two observables and, according to the third postulate, two Hermitian operators are associated with them.

*Momentum's eigenvalue problem*

Let us, for the beginning, consider that $\vec{p}$ is the particle's momentum, **P**$_x$ the operator for the momentum corresponding to the x-axis, $p_x$ its eigenvalue and $\Psi(x,y,z)$ the corresponding eigenfunction. The eigenvalue equation is:

$$\mathbf{P}_x \Psi(x,y,z) = p_x \Psi(x,y,z) \tag{35}$$

If we solve this equation, then we will know the eigenfunctions and eigenvalues, but, *first we must know the form of the operator* **P**$_x$. The third postulate does not explain how to find the operators corresponding to observables. So? The answer is similitude. One of the solutions of the differential equation of electromagnetic waves (and not only) is the plane wave. Let $\Psi(x,y,z)$ be such a monochromatic plane wave :

$$\psi(x,y,z) = ae^{i(\vec{k}\vec{r}-\omega t)} = ae^{i\varphi} \tag{36}$$

with "i" the imaginary number $i^2 = -1$, a = amplitude, $\vec{k}$ = the wave vector, $\vec{r}$ = position vector, $\omega$ = frequency, t = time, $\varphi$ = phase. Looking at the phase $\varphi$, we see that the derivative of phase with respect to coordinates gives the wave vector $\vec{k}$. According to de Broglie's hypothesis, a wave will be associated to each particle, and the wave vector is connected to the momentum, so the momentum operator can be associated to derivatives with respect to coordinates. Knowing that $E = h\nu = \hbar\omega$, $p = \dfrac{h}{\lambda} = \hbar k$, where h is Planck's constant (6.62607*10$^{-34}$ Js), $\hbar = \dfrac{h}{2\pi}$, $\omega = 2\pi\nu$ is the angular frequency (usually named frequency) and k is the wave number (modulus of the wave vector), $k = \dfrac{2\pi}{\lambda}$, $\Psi(x,y,z)$ can be written as:

$$\psi = ae^{\frac{i}{\hbar}(\vec{p}\vec{r}-Et)} = ae^{\frac{i}{\hbar}(xp_x+yp_y+zp_z-Et)} \tag{37}$$

Let's take the derivatives with respect to coordinates:

$$\frac{\partial \psi}{\partial x} = \frac{i}{\hbar}p_x\, ae^{\frac{i}{\hbar}(xp_x+yp_y+zp_z-Et+\hbar\alpha)} = \frac{i}{\hbar}p_x \psi$$

$$\frac{\partial \psi}{\partial y} = \frac{i}{\hbar}p_y\, ae^{\frac{i}{\hbar}(xp_x+yp_y+zp_z-Et+\hbar\alpha)} = \frac{i}{\hbar}p_y \psi \tag{38}$$

$$\frac{\partial \psi}{\partial z} = \frac{i}{\hbar}p_z\, ae^{\frac{i}{\hbar}(xp_x+yp_y+zp_z-Et+\hbar\alpha)} = \frac{i}{\hbar}p_z \psi$$

Writing:
$-i\hbar\dfrac{\partial}{\partial x}\Psi = p_x\Psi$, $-i\hbar\dfrac{\partial}{\partial y}\Psi = p_y\Psi$, $-i\hbar\dfrac{\partial}{\partial z}\Psi = p_z\Psi$ and comparing with

$\mathbf{P_x}\Psi = p_x\Psi$, $\mathbf{P_y}\Psi = p_y\Psi$, $\mathbf{P_z}\Psi = p_z\Psi$

we get:

$$\mathbf{P_x} = -i\hbar\frac{\partial}{\partial x}$$

$$\mathbf{P_y} = -i\hbar\frac{\partial}{\partial y} \tag{39}$$

$$\mathbf{P_z} = -i\hbar\frac{\partial}{\partial z}$$

Or, generally $\qquad\qquad\qquad \mathbf{P} = -i\hbar\nabla \tag{40}$

We have obtained the momentum operator, as derivatives with respect to coordinates.
Let $\Psi_{p_x}(x)$ be the wave-function corresponding to the $\mathbf{P_x}$ operator. Then, the eigenvalue equation for $\mathbf{P_x}$ is:

$$-i\hbar\frac{\partial \Psi_{p_x}(x)}{\partial x} = p_x \Psi_{p_x}(x) \tag{41}$$

Admitting momentum conservation and integrating equation (41), we will get:

$$\Psi_{p_x}(x) = Ne^{\frac{i}{\hbar}xp_x} \tag{42}$$

with N the norm constant.

What about eigenvalues? Momentum being the product of mass and velocity, the eigenvalue spectrum is continuous, because the velocity can take any value between -c and +c, (c light speed in vacuum), and mass can be any positive real number. Now, the momentum eigenfunction can be written as:

$$\Psi(x,p_x) = Ne^{\frac{i}{\hbar}xp_x} \tag{43}$$

because $p_x$ is a real variable like x. What about N? It will be determined very soon. The wavefunctions for $P_y$ and $P_z$ will be like (61) with corresponding coordinates and momenta.

*How does one norm eigenfunctions corresponding to continuous spectra of eigenvalues?*
Let us recall the direct product (12) and consider that we are dealing with a one-dimensional problem. It means that $\Phi$ and $\Psi$ can be written as:

$$\Phi(x) = \langle x|\Phi\rangle, \Psi(x) = \langle x|\Psi\rangle \tag{44}$$

and (12) becomes:

$$\langle \Phi|\Psi\rangle = \int_{all\,x} \langle \Phi|x\rangle\langle x|\Psi\rangle dx \tag{45}$$

Now, let us consider that $\Phi$ is just $|x'\rangle$, x and x′ being particle's positions on a straight line. (45) becomes:

$$\langle x'|\Psi\rangle = \int_{all\,x} \langle x'|x\rangle\langle x|\Psi\rangle dx \tag{46}$$

Recalling (44), we have:

$$\Psi(x') = \int_{all\,x} \langle x'|x\rangle \Psi(x) dx \tag{47}$$

An expression like (47) is valid if and only if $\langle x'|x\rangle$ is *Dirac's*, $\delta(x'-x)$, function. By definition, Dirac's function is:

$$\int_{-\infty}^{\infty} f(x)\delta(x-a)dx = f(a) \tag{48}$$

with $f(x) \to 0$ when $x \to \pm\infty$ and "a" a real number.

A usual function with such property does not exist. The integral, in the Riemann sense, of a function identical to zero excepting one point equals zero. By $\delta(x-a)$ we have to understand a limiting process, meaning that there are functions depending on a parameter $\varepsilon$, so that

$$\lim_{\varepsilon \to o} f(x,\varepsilon) = \delta(x) \tag{49}$$

*Important properties*:

$$\int_{-\infty}^{\infty} \delta(x)\, dx = 1 \tag{50}$$

With "a" a constant,
$$\delta(ax) = \frac{1}{|a|}\delta(x) \tag{51}$$

Fourier expansion,
$$\delta(x) = \frac{1}{2\pi}\int_{-\infty}^{\infty} e^{-i\alpha x}\, d\alpha \tag{52}$$

$$\delta(x) = \delta(-x) \tag{53}$$

So, if $\Psi(x,p_x)$ and $\Psi(x,p_x')$ are wavefunctions corresponding to the momentum continuous eigenvalue spectrum, then:

$$\int_{-\infty}^{\infty} \Psi^*(x,p_x)\,\Psi(x,p_x')\, dx = \delta(p_x - p_x') \tag{54}$$

Expression (54) is valid for any wavefunction corresponding to the continuous eigenvalue spectrum.

Using (43) in the above expression, we have:

$$N^*N \int_{-\infty}^{\infty} e^{\frac{i}{\hbar}x(p_x' - p_x)}\, dx = \delta(p_x' - p_x) \tag{55}$$

With properties (52) and (51), we have:

$$2\pi\hbar|N|^2\,\delta(p_x - p_x') = \delta(p_x - p_x') \tag{56}$$

Taking the integral of (56) and using the property (50) we get:

$$|N|^2 = \frac{1}{2\pi\hbar} \tag{57}$$

If N is real, then:

$$N = \frac{1}{\sqrt{2\pi\hbar}} \tag{58}$$

The momentum wavefunction will be:

$$\Psi(x,p_x) = \frac{1}{\sqrt{2\pi\hbar}}\, e^{\frac{i}{\hbar}xp_x} \tag{59}$$

and similarly
$$\Psi(y,p_y) = \frac{1}{\sqrt{2\pi\hbar}}\, e^{\frac{i}{\hbar}yp_y} \tag{60}$$

$$\Psi(z,p_z) = \frac{1}{\sqrt{2\pi\hbar}}\, e^{\frac{i}{\hbar}zp_z} \tag{61}$$

The general solution for **P** must be of the same type, namely an exponential of $\vec{r}\vec{p}$, and this can be done multiplying (59), (60), (61).

$$\Psi(x,y,z,p_x,p_y,p_z) = \frac{1}{(2\pi\hbar)^{3/2}} e^{\frac{i}{\hbar}\vec{r}\vec{p}} \qquad (62)$$

### *The position's eigenvalue problem*

To find the position operator we will use the similitude again. From Classical Mechanics [6] [32] we know the Poisson brackets. Let f and g be two functions of coordinates (x,y,z) and momenta ($p_x, p_y, p_z$). The Poisson bracket corresponding to these two functions is:

$$\{f,g\} = \sum_{j=x,y,z} \left( \frac{\partial f}{\partial p_j} \frac{\partial g}{\partial x_j} - \frac{\partial f}{\partial x_j} \frac{\partial g}{\partial p_j} \right) \qquad (63)$$

Taking $f = p_x$ and $g = x$, their Poisson bracket is:

$$\{p_x, x\} = 1 \qquad (64)$$

If we multiply the expression (64) by a function $\Psi(x)$, we have:

$$\{p_x, x\}\Psi = \Psi \qquad (65)$$

Now, let's take the derivative with respect to x of the product $x\Psi$.

$$\frac{\partial}{\partial x}(x\Psi) = \Psi + x\frac{\partial \Psi}{\partial x} \qquad (66)$$

Subtracting $x\frac{\partial \Psi}{\partial x}$ from both sides of (66), we get:

$$\frac{\partial}{\partial x}(x\Psi) - x\frac{\partial \Psi}{\partial x} = \Psi \qquad (67)$$

This expression (66), seems to be related to (65) if we can find something which is closed to Poisson bracket.

Now, let $\Psi(x)$ be the wavefunction of the position operator **X**.
The eigenvalue equation of **X** is:

$$\mathbf{X}\Psi = x\Psi \qquad (68)$$

Then, equation (67) can be written:

$$\frac{\partial}{\partial x}(\mathbf{X}\Psi) - \mathbf{X}\frac{\partial \Psi}{\partial x} = \Psi \quad \text{or} \quad \left(\frac{\partial}{\partial x}\mathbf{X} - \mathbf{X}\frac{\partial}{\partial x}\right)\Psi = \Psi \qquad (69)$$

The transition from (67) to (69) can be made only if the action of the ***position operator*** consists in ***multiplying*** a function. In the expression (69) the bracket seems to play a similar role with the Poisson bracket. Has it any significance? Yes, it is also an operator, more precisely the commutator of the operators **X** and $\frac{\partial}{\partial x}$, and perhaps Dirac followed a similar way to introduce the operators' commutativity and the corresponding properties. The eigenvalues of the position operator make up a continuous spectrum because the position of a particle can be any real number.

*The angular momentum's eigenvalue problem*

Classical Mechanics told us that there is a physical quantity called angular momentum, which is a vector, defined as:

$$\vec{L} = L_x \vec{i} + L_y \vec{j} + L_z \vec{k} = \vec{r} \times \vec{p} = \begin{vmatrix} \vec{i} & \vec{j} & \vec{k} \\ x & y & z \\ p_x & p_y & p_z \end{vmatrix} \quad (70)$$

where $\vec{r}$ is the particle's vector position with respect to a reference system whose basis are the unit vectors $\vec{i}, \vec{j}, \vec{k}$, and $\vec{p}$ is the particle's momentum. From (70) we get:

$$L_x = y p_z - z p_y$$
$$L_y = z p_x - x p_z \quad (71)$$
$$L_z = x p_y - y p_x$$

Knowing that to the observables position and momentum correspond Hermitian operators **X**, **Y**, **Z**, and **P_x**, **P_y**, **P_z** respectively, we can get the angular momentum operators, by replacing the classical observables by corresponding operators according to the third postulate.

$$\mathbf{L_x} = \mathbf{YP_z} - \mathbf{ZP_y} = i\hbar \left( z \frac{\partial}{\partial y} - y \frac{\partial}{\partial z} \right)$$

$$\mathbf{L_y} = \mathbf{ZP_x} - \mathbf{XP_z} = i\hbar \left( x \frac{\partial}{\partial z} - z \frac{\partial}{\partial x} \right) \quad (72)$$

$$\mathbf{L_z} = \mathbf{XP_y} - \mathbf{YP_x} = i\hbar \left( y \frac{\partial}{\partial x} - x \frac{\partial}{\partial y} \right)$$

*Note*. In (72) "i" is the imaginary number $i^2 = -1$

Also, $\quad\quad\quad\quad\quad\quad \mathbf{L}^2 = \mathbf{L_x^2} + \mathbf{L_y^2} + \mathbf{L_z^2} \quad (73)$

To find the angular momentum's operators it is easier to pass to spherical system of coordinates. Let $f(x,y,z)$ be a certain function of coordinates. In spherical coordinates we have:

$$x = r \sin\vartheta \cos\varphi$$
$$y = r \sin\vartheta \sin\varphi \quad (73)$$
$$z = r \cos\vartheta$$

In (72) we have to replace the derivatives with respect to x, y, z with $r, \vartheta, \varphi$ derivatives. Connection $\left( \frac{\partial}{\partial r}, \frac{\partial}{\partial \vartheta}, \frac{\partial}{\partial \varphi} \right) \to \left( \frac{\partial}{\partial x}, \frac{\partial}{\partial y}, \frac{\partial}{\partial z} \right)$ will be found solving the following system of algebraic equation with respect to the unknowns $\left( \frac{\partial f}{\partial x}, \frac{\partial f}{\partial y}, \frac{\partial f}{\partial z} \right)$:

$$\frac{\partial f}{\partial r} = \sin\vartheta\cos\varphi\frac{\partial f}{\partial x} + \sin\vartheta\sin\varphi\frac{\partial f}{\partial y} + \cos\vartheta\frac{\partial f}{\partial z}$$

$$\frac{\partial f}{\partial \vartheta} = -r\cos\vartheta\cos\varphi\frac{\partial f}{\partial x} + r\cos\vartheta\sin\varphi\frac{\partial f}{\partial y} - r\sin\vartheta\frac{\partial f}{\partial z} \quad (74)$$

$$\frac{\partial f}{\partial \varphi} = -r\sin\vartheta\sin\varphi\frac{\partial f}{\partial x} + r\sin\vartheta\cos\varphi\frac{\partial f}{\partial y}$$

After some calculations, we get:

$$\mathbf{L_x} = -i\hbar\left(\sin\varphi\frac{\partial}{\partial \vartheta} + \cos\varphi\, \text{ctan}\vartheta\frac{\partial}{\partial \varphi}\right)$$

$$\mathbf{L_y} = -i\hbar\left(\cos\varphi\frac{\partial}{\partial \vartheta} - \cos\varphi\, \text{ctan}\vartheta\frac{\partial}{\partial \varphi}\right) \quad (75)$$

$$\mathbf{L_z} = -i\hbar\frac{\partial}{\partial \varphi}$$

With (75) and (73) we, also, will get:

$$\mathbf{L}^2 = -\hbar^2\left(\frac{1}{\sin\vartheta}\frac{\partial}{\partial \vartheta}(\sin\vartheta\frac{\partial}{\partial \vartheta}) + \frac{1}{\sin^2\vartheta}\frac{\partial^2}{\partial \varphi^2}\right) \quad (76)$$

An important result we have got: *the angular momentum's operators* act upon the *angular coordinates* $\vartheta$ and $\varphi$ only.

*$L_z$ eigenvalue problem*

Because $\mathbf{L_z}$ acts upon the variable $\varphi$ only, let's consider the corresponding eigenfunctions $\Phi(\varphi)$. The $\mathbf{L_z}$'s eigenvalue equation will be:

$$\mathbf{L_z}\Phi(\varphi) = \mu\Phi(\varphi) \quad (77)$$

With $\mathbf{L_z}$ done by (75) the eigenvalue equation takes the form:

$$-i\hbar\frac{d\Phi(\varphi)}{d\varphi} = \mu\Phi(\varphi) \quad (78)$$

(78) is a simple differential equation with separable variables, and recalling the angular momentum conservation law (μ constant), the solution is:

$$\Phi(\varphi) = Ne^{\frac{i}{\hbar}\mu\varphi} \quad (79)$$

Now, we have to determine the eigenvalues μ and the norm constant N. The uniformity condition of the wave-functions claims

$$\Phi(\varphi) = \Phi(\varphi + 2\pi) \quad (80)$$

which gives:

$$\mu = m\hbar, \quad \text{with } m = 0, \pm 1, \pm 2, \ldots \quad (81)$$

So, the *eigenvalue spectrum* of the *angular momentum* operator $\mathbf{L_z}$ is a *discrete* one: *the eigenvalue can be* (in principle) *any whole multiple of* $\hbar$, *including zero* too.

Because the eigenvalue spectrum is discrete, the norm condition: $\int_0^{2\pi} |\Phi_m(\varphi)|^2 \, d\varphi = \int_0^{2\pi} N^* N e^{-im\varphi} e^{im\varphi} d\varphi = 1$ gives (if N is a real constant):

$N = \dfrac{1}{\sqrt{2\pi}}$. So, the angular momentum's $\mathbf{L_z}$ eigenfunction is:

$$\Phi_m(\varphi) = \frac{1}{\sqrt{2\pi}} e^{im\varphi} \tag{82}$$

What is the *significance of* "m"? We will find out soon.

## $L^2$ eigenvalue problem
Let's have:

$$\mathbf{\Omega} = \frac{1}{\sin\vartheta} \frac{\partial}{\partial\vartheta}\left(\sin\vartheta \frac{\partial}{\partial\vartheta}\right) + \frac{1}{\sin^2\vartheta} \frac{\partial^2}{\partial\varphi^2} \tag{83}$$

Then: 
$$\mathbf{L}^2 = -\hbar^2 \mathbf{\Omega} \tag{84}$$

Because $\mathbf{L}^2$ acts upon the angular variables only, let $Y(\vartheta,\varphi)$ be its eigenfunction. The eigenvalue equation will be:

$$\mathbf{L}^2 Y(\vartheta,\varphi) = \omega Y(\vartheta,\varphi) \quad \text{or} \quad \mathbf{\Omega} Y(\vartheta,\varphi) + \frac{\omega}{\hbar^2} Y(\vartheta,\varphi) = 0 \tag{85}$$

The equation (85) is known as the spherical waves' differential equation being solved in any book of "Special Functions", see, e.g. [8] or [5]. The solutions of (85) are the spherical functions $Y_{lm}(\vartheta,\varphi)$:

$$Y_{lm}(\vartheta,\varphi) = P_{lm}(\cos\vartheta)\Phi_m(\varphi) \tag{86}$$

$P_{lm}(\cos\vartheta)$ being the associated Legendre polynomials of the first kind:

$$P_{lm}(\cos\vartheta) = (-1)^l \sqrt{\frac{2l+1}{2}\frac{(l+m)!}{(l-m)!}} \frac{1}{2^l l!(1-\cos^2\vartheta)^{m/2}} \cdot \frac{d^{l-m}}{(d\cos\vartheta)^{l-m}}(1-\cos^2\vartheta)^l \tag{87}$$

With (82) we have:

$$Y_{lm}(\vartheta,\varphi) = (-1)^l \sqrt{\frac{2l+1}{2}\frac{(l+m)!}{(l-m)!}} \frac{1}{2^l l!(1-\cos^2\vartheta)^{m/2}} \cdot \frac{d^{l-m}}{(d\cos\vartheta)^{l-m}}(1-\cos^2\vartheta)^l \frac{1}{\sqrt{2\pi}} e^{im\varphi} \tag{88}$$

From (88) we see that for $|m| > l$, $Y_{lm}(\vartheta,\varphi)$ vanish, so, there is no solution of (85). Hence "m", for fixed $l$ can take the values:

$$m = -l, -l+1, \ldots -1, 0, 1, \ldots l-1, l \tag{89}$$

The $\mathbf{L}^2$ eigenvalues are $\hbar^2 l(l+1)$ with $l = 0, 1, 2, 3, \ldots$ \hfill (90)

What are these numbers *l* and *m* ? As one can see, *l* and *m* are connected to the eigenvalues of the angular momentum operators $\mathbf{L}^2$ and $\mathbf{L_z}$. They are called **quantum numbers**: *l* is the *orbital quantum number* and *m* is the *magnetic quantum number.* Why are these numbers called *orbital* and *magnetic* respectively? I think there are historical reasons: the first application of Quantum Mechanics was the simplest atom, namely, hydrogen, so, the electron orbiting the nucleus has an angular momentum $\vec{L}$, and $L_z$ its projection onto the z-axis (an external magnetic field, e.g.), hence we have got *l* as *orbital quantum number* and *m* as *magnetic quantum number* which has nothing magnetic in itself.

**Dynamics**
## 4. The forth postulate of Quantum Mechanics

*The time evolution of a quantum state is governed by a unitary transformation.*
*If $\Psi(t)$ is the probability amplitude of a quantum state at time $t$, then $\Psi(t+\Delta t)$ is its probability amplitude at a later time $t+\Delta t$, so that $|\Psi(t+\Delta t)\rangle = \mathbf{U}(t+\Delta t, t)|\Psi(t)\rangle$, where $\mathbf{U}$ is a unitary linear operator.*

$\mathbf{U}$ is unitary if $\mathbf{U}\mathbf{U}^+ = \mathbf{U}^+\mathbf{U} = 1$, where $\mathbf{U}^+$ is $\mathbf{U}$'s adjoint operator.
To find out something about $\mathbf{U}$ we will proceed in a similar way with Feynman's [2], and let us consider $\Psi_n$ the basis vectors, so the expression $|\Psi(t+\Delta t)\rangle = \mathbf{U}(t+\Delta t, t)|\Psi(t)\rangle$ can be written as:

$$\langle \Psi_n | \Psi(t+\Delta t)\rangle = \sum_k \langle \Psi_n | \mathbf{U}(t+\Delta t, t) | \Psi_k \rangle \langle \Psi_k | \Psi(t)\rangle \tag{91}$$

With $\quad\quad\quad\quad \langle \Psi_n | \mathbf{U}(t+\Delta t, t) | \Psi_k \rangle = U_{nk}(t+\Delta t, t) \tag{92}$

(91) becomes:
$$\langle \Psi_n | \Psi(t+\Delta t)\rangle = \sum_k U_{nk}(t+\Delta t, t) \langle \Psi_k | \Psi(t)\rangle \tag{93}$$

What do we know about the matrix $U_{nk}(t+\Delta t, t)$?

Well, first of all, when $\Delta t \to 0$ we have to reach the initial state, which means that $U_{nn}(t,t) \to 1$ and $U_{nk}(t,t) \to 0, n \neq k$, therefore $U_{nk}$ can be written as:

$$U_{nk} = \delta_{nk} - \frac{i}{\hbar} H_{nk} \Delta t \tag{94}$$

Indeed, (94) (constants "i" and $\hbar$ are introduced for convenience) satisfy the above conditions, when $\Delta t \to 0$, $\delta_{nk}$ being the Kronecker symbol and $H_{nk}$ another matrix whose significance will be clarified soon. With (94), (93) becomes:

$$\langle \Psi_n | \Psi(t+\Delta t)\rangle = \langle \Psi_n | \Psi(t)\rangle - \frac{i}{\hbar} \Delta t \sum_k H_{nk} \langle \Psi_k | \Psi(t)\rangle \tag{95}$$

Or:

$$\langle\Psi_n|\Psi(t+\Delta t)\rangle - \langle\Psi_n|\Psi(t)\rangle = -\frac{i}{\hbar}\Delta t \sum_k H_{nk}\langle\Psi_k|\Psi(t)\rangle \Rightarrow$$

$$\frac{\langle\Psi_n|\Psi(t+\Delta t)\rangle - \langle\Psi_n|\Psi(t)\rangle}{\Delta t} = -\frac{i}{\hbar}\sum_k \langle\Psi_n|\mathbf{H}|\Psi_k\rangle\langle\Psi_k|\Psi(t)\rangle$$

$$\frac{\langle\Psi_n|\Psi(t+\Delta t) - \Psi(t)\rangle}{\Delta t} = -\frac{i}{\hbar}\sum_k \langle\Psi_n|\mathbf{H}|\Psi_k\rangle\langle\Psi_k|\Psi(t)\rangle \Rightarrow$$

$$\frac{|\Psi(t+\Delta t) - \Psi(t)\rangle}{\Delta t} = -\frac{i}{\hbar}\mathbf{H}\sum_k |\Psi_k\rangle\langle\Psi_k|\Psi(t)\rangle$$

Passing the limit $\Delta t \to 0$ in the above expression, we get:

$$i\hbar\frac{d|\Psi(t)\rangle}{dt} = \mathbf{H}|\Psi(t)\rangle \qquad (96)$$

In the above equation **H** is going to play a major role. What is it? Let's find out! Firstly, if **H** doesn't depend on time, starting from (96), we will get $\Psi(t) \propto e^{-\frac{i}{\hbar}\text{something}\, t}$.

What can be that "something"? Let's come back to classical Physics..
The expression (37) is a plane wave, and it seems to be the "relative" of the solution containing "something". By comparison we can think of "something" as being energy. Taking the time derivative of (37), we get:

$$\frac{\partial\Psi(t)}{\partial t} = -\frac{i}{\hbar}E\Psi(t) \Rightarrow i\hbar\frac{\partial\Psi(t)}{\partial t} = E\Psi(t) \qquad (96')$$

From (95) and (96) we conclude that **H** must be *the **operator** corresponding **to the total energy*** of the physical system.

What about **U**? Expanding $\Psi(t+\Delta t)$ in Taylor series around t, we have:

$$\Psi(t+\Delta t) = \Psi(t) + \Delta t\, \partial_t \Psi(t) + \frac{1}{2!}(\Delta t)^2 \partial_t^2 \Psi(t) + \ldots =$$

$$= (1 + \Delta t\, \partial_t + \frac{1}{2!}(\Delta t)^2 \partial_t^2 + \cdots)\Psi(t) =$$

$$= \left(\sum_{k=0} \frac{1}{k!}(\Delta t)^k \partial_t^k\right)\Psi(t) = e^{\Delta t\, \partial_t}\Psi(t)$$

For convenience, the above expression can be written as:

$$\Psi(t+\Delta t) = e^{i\mathbf{S}}\Psi(t), \text{ which gives } \mathbf{U} = e^{i\mathbf{S}} \qquad (97)$$

Here **S** is a Hermitian operator. From (97) one can see that the unitarity condition is fulfilled.

*The Schrödinger equation*
Classical Physics tells us that the total energy of a physical system is the sum of the kinetic energy and potential energy:

$$E = \frac{m_0 v^2}{2} + E_p = \frac{p^2}{2m_0} + E_p \qquad (98)$$

Obviously, in the above expression, the total energy is the non-relativistic energy.
*Note*
Here, and everywhere else (unless it is explicitly said), by "m" we will understand the particle's (or physical system's) mass in the sense of Einstein's Theory of Relativity, namely, the particle's mass in the reference system where the particle (or the physical system) is at rest.
In (98) we see two kinds of physical quantities, or (if measurable), two kinds of observables: momentum, whose self-adjoint operators are derivatives with respect to coordinates, and potential energy. The momentum has its own operators, but what about the potential energy? The potential energy depends on the particle's (or physical system's) coordinates, so it is not that "crazy" to think that the operator corresponding to the potential energy is a multiplicative one, much like the position's operator. Hence, the total energy's operator will be found by replacing the classical quantities with quantum corresponding operators.

$$\mathbf{H} = -\frac{\hbar^2}{2m_0}\nabla^2 + \mathbf{U} = -\frac{\hbar^2}{2m_0}\Delta + \mathbf{U} \qquad (99)$$

Where $\nabla^2 = \Delta = \frac{\partial^2}{\partial x^2} + \frac{\partial^2}{\partial y^2} + \frac{\partial^2}{\partial z^2}$ is Laplace's operator, or Laplacean, and **U** is the multiplicative operator corresponding to potential energy. From Classical Physics [1,9], we know that the total energy of a physical system is just Hamilton's function, hence **H** will be called *Hamiltonian.* **H** may be, or may not be dependent on time. With **H** given by (99), equation (96) becomes:

$$i\hbar \frac{d\Psi(x,y,z,t)}{dt} = -\frac{\hbar^2}{2m_0}\Delta\Psi(x,y,z,t) + \qquad (100)$$
$$+ U(x,y,z)\Psi(x,y,z,t)$$

This is the *fundamental equation of Quantum Mechanics* and is called the **Schrödinger** equation.
Can we split the function $\Psi(x,y,z,t)$ into a function depending on coordinates only, and the other depending on time? The plane wave (again!) and equation (96) suggest that we can write:

$$\Psi(x,y,z,t) = \Psi(x,y,z) \cdot e^{-\frac{i}{\hbar}Et} \qquad (101)$$

Introducing (101) in (100) we have:

$$-\frac{\hbar^2}{2m}\Delta\Psi(x,y,z) + U(x,y,z)\Psi(x,y,z) = E\Psi(x,y,z) \qquad (102)$$

Equation (102) is, also, the **Schrödinger** equation, but, this version of the equation is independent of time. In other words, it is *the stationary Schrödinger* equation. This equation (102) is the *energy eigenvalue* equation, which tells us that if the physical system has the energy E at the initial time, then at any subsequent time it will have the same energy. The *stationary Schrödinger* equation is the fundamental equation of the microscopic world with definite energies, e.g. molecules, atoms, nuclei, etc. Knowing the potential energy, or the interaction potential U(x,y,z), by integrating the stationary Schrödinger equation we can find the energy eigenvalues and eigenfunctions of the physical system.

**Observation**. *It is necessary to outline that the operator* **H** *in the time-dependent Schrödinger* equation (100) is the *Hamiltonian of the system,* and **H** *from* the *stationary Schrödinger* equation is the *operator energy* . The two **H** are identical only if the Hamiltonian doesn't depend on time.

*An important property*

If $\Psi$ and $\Phi$ are two solutions of the time dependent Schrödinger equation, then their inner product is constant. The Schrödinger equations for these functions and the inner product are:

$$\mathbf{H}\Phi = i\hbar \frac{\partial \Phi}{\partial t}, \quad \mathbf{H}\Psi = i\hbar \frac{\partial \Psi}{\partial t}, \quad \langle \Phi | \Psi \rangle$$

Taking the time derivative of the inner product and making use of the fact that $\Psi$ and $\Phi$ satisfy the time dependent Schrödinger equations, and also, that **H** is the Hermitian we will have:

$$\frac{d\langle \Phi | \Psi \rangle}{dt} = \left\langle \frac{d\Phi}{dt} \middle| \Psi \right\rangle + \left\langle \Phi \middle| \frac{d\Psi}{dt} \right\rangle = \left\langle \frac{1}{i\hbar} \mathbf{H}\Phi \middle| \Psi \right\rangle +$$

$$\left\langle \Phi \middle| \frac{1}{i\hbar} \mathbf{H}\Psi \right\rangle = -\frac{1}{i\hbar} \langle \mathbf{H}\Phi | \Psi \rangle + \frac{1}{i\hbar} \langle \Phi | \mathbf{H}\Psi \rangle =$$

$$\frac{1}{i\hbar} \left( \langle \Phi | \mathbf{H}\Psi \rangle - \langle \mathbf{H}\Phi | \Psi \rangle \right) = \frac{1}{i\hbar} \left( \langle \Phi | \mathbf{H}\Psi \rangle - \langle \Phi | \mathbf{H}\Psi \rangle \right) = 0$$

The result is: $\qquad \langle \Phi | \Psi \rangle = \text{constant}$ (103)

If $\qquad \Phi \equiv \Psi, \langle \Psi | \Psi \rangle_{\text{initial time}} = \langle \Psi | \Psi \rangle_{\text{any subsequent time}}$ (104)

**Measurements**
**5. The fifth postulate of Quantum Mechanics**

The fifth postulate outlines the statistical nature of Quantum Mechanics and bridges the mathematical apparatus introduced by the first four postulates and the experimental results of a measuring process.

Let $\mathcal{F}$ be an observable and **F** the corresponding Hermitian operator. With $\Psi$(x,y,z) the eigenfunctions, the eigenvalue equation can be written as:

$$\mathbf{F}\Psi = f\Psi \qquad (105)$$

Considering the general case of a joint (discrete + continuous) eigenvalue spectrum, the superposition principle may be written as:

$$\Psi(x,y,z) = \sum_{k=1}^{n} c_k \Psi_k(x,y,z) + \int_{all\,\alpha} c(\alpha)\Psi(x,y,z,\alpha)d\alpha \qquad (106)$$

where $\Psi_k$ (k = 1,2,3….) are the eigenfunctions corresponding to the discrete spectrum, and $\Psi(x,y,z,\alpha) \equiv \Psi_\alpha(x,y,z)$ the eigenfunctions corresponding to the continuous spectrum (α any real number). We already know that the expansion coefficients are:

$$\begin{aligned} c_k &= \langle \Psi_k | \Psi \rangle \\ c(\alpha) &= \langle \Psi_\alpha | \Psi \rangle \end{aligned} \qquad (107)$$

Question. If we arrange an experimental set-up to measure the observable $\mathcal{F}$, then what are the values we will expect to get? The answer is given by the *fifth postulate*:

*As a result of a measuring process performed upon an observable $\mathcal{F}$, we will obtain only the eigenvalues of the Hermitian operator, **F**, associated to the observable. The probability of getting an eigenvalue $f_k$ corresponding to the discrete spectrum is $|c_k|^2$, and the probability of getting an eigenvalue $f_\alpha$ corresponding to the continuous spectrum within an interval $d\alpha$ is $|c_\alpha|^2 d\alpha$.*

Over the years, the measuring process in Quantum Mechanics gave rise to a lot of discussions, not only concerning physics, but also philosophy. As we have seen before, there are *entangled* states, so if we have two separate particles being (each one of them) in definite states, then the system of both particles is in an indefinite state. Furthermore, if we measure a physical quantity for one particle, then Quantum Mechanics predicts the correct value of the same physical quantity for the second particle *without* measurements, as we shall see soon.

*The expected value of an observable*

Let "a" be a random variable taking (along a measuring process) the values $a_1, a_2, \ldots \ldots a_n$. By definition its expected value $\langle a \rangle$ is:

$$\langle a \rangle = \sum_k a_k P_k \qquad (108)$$

where $P_k$ is the probability of getting the value $a_k$.

Now, let us choose an observable $\mathcal{F}$ whose corresponding operator, according to the third postulate, is **F**. The fifth postulate says that during the measuring process we will get the eigenvalues "f" of this operator only. Let, also, $\Psi_k(x,y,z)$ be the eigenfunctions corresponding to the discrete spectrum and $\Psi(x,y,z,\alpha)$ the eigenfunctions from the continuous spectrum. The eigenvalue equation is (102). Using formula (108) and the "words" of the fifth postulate, the expected value of $\mathcal{F}$ will be:

$$<\mathcal{F}> = \sum_k f_k |c_k|^2 + \int_\alpha f |c(\alpha)|^2 d\alpha \qquad (109)$$

The complex conjugates of (107) being $\begin{aligned} c_k^* &= \langle \Psi | \Psi_k \rangle \\ c^*(\alpha) &= \langle \Psi | \Psi_\alpha \rangle \end{aligned}$, we will get:

$$\sum_k f_k c_k \langle \Psi | \Psi_k \rangle + \int_\alpha f c(\alpha) \langle \Psi | \Psi_\alpha \rangle d\alpha =$$

$$\sum_k c_k \langle \Psi | f_k \Psi_k \rangle + \int_\alpha c(\alpha) \langle \Psi | f \Psi_\alpha \rangle d\alpha =$$

$$\sum_k c_k \langle \Psi | \mathbf{F} \Psi_k \rangle + \int_\alpha c(\alpha) \langle \Psi | \mathbf{F} \Psi_\alpha \rangle d\alpha =$$

$$\left\langle \Psi \left| \mathbf{F} \left( \sum_k c_k \Psi_k \right) \right\rangle + \left\langle \Psi \left| \mathbf{F} \left( \int_\alpha c(\alpha) \Psi_\alpha d\alpha \right) \right\rangle \right. =$$

$$\left\langle \Psi \left| \mathbf{F} \left( \sum_k c_k \Psi_k + \int_\alpha c(\alpha) \Psi_\alpha d\alpha \right) \right\rangle \right. = \langle \Psi | \mathbf{F} \Psi \rangle$$

So, the expected value of $\mathcal{F}$ is an inner product:

$$<\mathcal{F}> = \langle \Psi | \mathbf{F} \Psi \rangle \qquad (110)$$

***The time derivative of the expected value***
Let's take the time derivative of the expression (110).

$$\frac{d}{dt} <\mathcal{F}> = \left\langle \frac{\partial \Psi}{\partial t} \middle| \mathbf{F} \Psi \right\rangle + \left\langle \Psi \middle| \frac{\partial \mathbf{F}}{\partial t} \Psi \right\rangle + \left\langle \Psi \middle| \mathbf{F} \frac{\partial \Psi}{\partial t} \right\rangle$$

Since $\psi$ is a solution of the Schrödinger equation (100), we have:

$$\frac{d}{dt} <\mathcal{F}> = \left\langle -\frac{i}{\hbar} \mathbf{H} \Psi \middle| \mathbf{F} \Psi \right\rangle + \left\langle \Psi \middle| \frac{\partial \mathbf{F}}{\partial t} \Psi \right\rangle + \left\langle \Psi \middle| -\frac{i}{\hbar} \mathbf{F} \mathbf{H} \Psi \right\rangle$$

Or, because **H** is a Hermitian operator:

$$\frac{d}{dt} <\mathcal{F}> = \left\langle \Psi \middle| \frac{i}{\hbar} \mathbf{H} \mathbf{F} \Psi \right\rangle + \left\langle \Psi \middle| -\frac{i}{\hbar} \mathbf{F} \mathbf{H} \Psi \right\rangle + \left\langle \Psi \middle| \frac{\partial \mathbf{F}}{\partial t} \Psi \right\rangle$$

$$\frac{d}{dt} <\mathcal{F}> = \left\langle \Psi \middle| \left( \frac{\partial \mathbf{F}}{\partial t} + \frac{i}{\hbar} (\mathbf{H} \mathbf{F} - \mathbf{F} \mathbf{H}) \right) \Psi \right\rangle \qquad (111)$$

If the operator **F** doesn't explicitly depend on time, we have:

$$\frac{d}{dt} <\mathcal{F}> = \left\langle \Psi \middle| \frac{i}{\hbar} (\mathbf{H} \mathbf{F} - \mathbf{F} \mathbf{H}) \Psi \right\rangle = \left\langle \Psi \middle| \frac{i}{\hbar} [\mathbf{H}, \mathbf{F}] \Psi \right\rangle \qquad (112)$$

A very interesting expression! If the operators **H** and **F** commute, then the time derivative of the observable $\mathcal{F}$ is zero, which means the observable remains constant.

Here is a symmetry statement: *for any physical system with the Hamiltonian* **H**, *the observables which will be conserved are those whose operators commute with* **H**. **And more**: *if two operators commute, they have simultaneous eigenfunctions.* For details, see e.g.[5]. For example, the operators **X** and **P_y** commute, which means that the observable "x" and "$p_y$" are independent variables and can be precisely measured simultaneously.

**And even more**: *the commutativity of operators outlines the possibility of measuring simultaneously the corresponding observables.* For instance, operators **X** and **P_x**, and the next two pairs, do not commute,

$$[\mathbf{X},\mathbf{P}_x] = [\mathbf{Y},\mathbf{P}_y] = [\mathbf{Z},\mathbf{P}_z] = i\hbar \tag{113}$$

and it means that the corresponding pairs of variables $(x, p_x), (y, p_y), (z, p_z)$ can not be precisely measured simultaneously, as the uncertainty principle says.

*Uncertainty relations*

We have learned enough to deduce the general form of the uncertainty relations.
Let us consider two Hermitian operators **F** and **G**. Let **C** be the commutator of the two:
$$\mathbf{C} = [\mathbf{F},\mathbf{G}] = \mathbf{FG} - \mathbf{GF}$$
The operator i**C** is also Hermitian, as one can easily prove.
Lets also, consider **Q** = **F** + i α **G**, with α a real constant. Because the adjoint operator of **Q** is **Q**$^+$ = **F** - i α **G**, **Q** is not Hermitian. Let's have **R** = **Q**$^+$**Q** = (**F** - iα**G**)( **F** + iα**G**) = **F**$^2$ + α i**C** + α$^2$ **G**$^2$,
and $\Re$ be the **R**'s eigenvalue. Then, if Ψ is the probability amplitude describing the quantum system, according to (129), the expected value of $\Re$ is:

$$<\Re> = \langle\Psi|\mathbf{Q}^+\mathbf{Q}\Psi\rangle = \langle\mathbf{Q}\Psi|\mathbf{Q}\Psi\rangle \geq 0$$

$<\Re> = 0$, if and only if **Q**Ψ = 0. On the other side,
$$<\Re> = <\mathcal{F}^2 + \alpha i \mathcal{C} + \alpha^2\mathcal{G}^2> = <\mathcal{F}^2> + \alpha<i\mathcal{C}> + \alpha^2<\mathcal{G}^2> \geq 0$$

The last expression is an algebraic equation of the second degree, coefficients $<\mathcal{F}^2>$, $<i\mathcal{C}>, <\mathcal{G}^2>$ being real numbers because the corresponding operators are Hermitian. So, in order to be satisfied, one has to have: $<i^2\mathcal{C}^2> - 4 <\mathcal{F}^2><\mathcal{G}^2> \leq 0$, or

$$<\mathcal{F}^2><\mathcal{G}^2> \geq \frac{1}{4}\left|\langle\mathcal{C}^2\rangle\right|$$

Equal ( =) sign means the root of the above algebraic equation is double and real, and so, **Q**Ψ = 0.
Let $\mathcal{A}$ and $\mathcal{B}$ be two observables and **A** and **B** the corresponding Hermitian operators. We can choose $\mathcal{F} = \mathcal{A} - <\mathcal{A}>$ and $\mathcal{G} = \mathcal{B} - <\mathcal{B}>$, then $<\mathcal{F}^2> = <\Delta\mathcal{A}^2>$ and $<\mathcal{G}^2> = <\Delta\mathcal{B}^2>$, so we have: $<\Delta\mathcal{A}^2><\Delta\mathcal{B}^2> \geq \frac{1}{4}\left|\langle\mathcal{C}^2\rangle\right|$. If $\Delta\mathcal{A} = (<\Delta\mathcal{A}^2>)^{1/2}$ and $\Delta\mathcal{B} = (<\Delta\mathcal{B}^2>)^{1/2}$, assimilate with standard deviations, then we will get the most general form of the uncertainty relations:

$$\Delta\mathcal{A}\,\Delta\mathcal{B} \geq \frac{1}{2}\left|\langle\mathcal{C}\rangle\right| \tag{114}$$

$\mathcal{C}$ is the observable corresponding to the commutator operator **C**, and $\mathcal{C}$ becomes zero only if **C** = 0, which means that **F** and **G** commute, and $\Delta\mathcal{A}\,\Delta\mathcal{B} \geq 0$, equal sign (=) meaning that $\Delta\mathcal{A}$ may be zero, or $\Delta\mathcal{B}$ can be zero, or even both of them are null. Such a situation opens up the possibility of measuring exactly, and simultaneously, two physical quantities, the corresponding operators **A** and **B**, for the same wave function, Ψ.

If $C \neq 0$ there is no possibility of having both $\Delta \mathcal{A} = 0$ and $\Delta \mathcal{B} = 0$, which means that, by no means, can we have a quantum physical system that allows to get well determined values simultaneously. If, for instance the observable $\mathcal{A}$ is precisely determined, then $\Delta \mathcal{A} = 0$, without having any information about $\mathcal{B}$ because $\Delta \mathcal{B} \to \infty$, and reversely.

If we take **F** = **X** and **G** = **P$_x$**, then **C** = [**X**, **P$_x$**] = $i\hbar$, $|C| = \hbar$, $\langle |C| \rangle = \hbar$, and also, $\Delta \mathcal{A} = \Delta x$, $\Delta \mathcal{B} = \Delta p_x$, according to Heisenberg's formula $\Delta x \, \Delta p_x \geq \dfrac{\hbar}{2}$.

*A couple of things that will prove themselves to be useful one day*

Let's have a physical system described by the amplitude probability Ψ(x,y,z,t) and let $\rho_\Psi = |\Psi|^2$ be the probability density. $\rho_\Psi$ has the dimensions of m$^{-3}$, so it can be used in finding the density of a number of particles, the charge density, the mass density, and so on.

a) If N is the total number of particles, $\rho_\Psi$ N will be the density of the particles.
b) If e is the electron's charge, and Ne is the total electric charge carried, $\rho_\Psi$ Ne is the charge density.
c) If m$_0$ is the mass of a particle, and N is the total number of particles, then $\rho_\Psi$ N m$_0$ is the mass density.

If Ψ(x,y,z,t) satisfies the Schrödinger equation, then its complex conjugate Ψ*(x,y,z,t) will be also a solution. Let's do some calculations. The Schrödinger equations for Ψ(x,y,z,t) and Ψ*(x,y,z,t) multiplied by Ψ*(x,y,z,t) and Ψ(x,y,z,t) respectively, are:

$$\Psi^* i\hbar \frac{d\Psi}{dt} = -\frac{\hbar^2}{2m_0} \Psi^* \Delta \Psi + U \Psi^* \Psi$$

$$-\Psi i\hbar \frac{d\Psi^*}{dt} = -\frac{\hbar^2}{2m_0} \Psi \Delta \Psi^* + U \Psi \Psi^*$$

Subtracting, part by part, the two equations above, and knowing that

$$\vec{j}_\Psi = \frac{i\hbar}{2m_0} \left( \Psi \nabla \Psi^* - \Psi^* \nabla \Psi \right) \quad (115)$$

we will have, after some calculations;

$$\frac{\partial \rho_\Psi}{\partial t} + \nabla \vec{j}_\Psi = 0 \quad (116)$$

Equation (116) is nothing else but the *continuity equation*, $\vec{j}_\Psi$ *being the probability density current*. With (116) we have different conservation laws. Let's integrate this equation on some volume V encompassed by a surface Σ:

$$\int_V \frac{\partial \rho_\Psi}{\partial t} dV = -\int_V \nabla \vec{j}_\Psi dV = -\oint_\Sigma \vec{j}_\Psi \vec{n} dA$$

Or,
$$\frac{\partial}{\partial t}\int_V \rho_\Psi dV = -\oint_\Sigma \vec{j}_\Psi \vec{n} dA \tag{117}$$

In the above calculations we made use of the Gauss formula connecting volume and surface integrals, $\vec{n}$ being the outer normal of the elementary surface dA. Reading formula (117), we can say that the change in time of the probability equals the negative of the flux of the probability current density through the surface $\Sigma$. If we integrate over the entire space, then $\Sigma$ becomes the surface tending to infinity and taking into account the properties of the vectors of Hilbert space (the wave-functions at infinity are zero) the right side of the expression (117) will cancel, conserving probability and other quantities constructed with $\rho_\Psi$.

*The classical limit of Quantum Mechanics*

There is a principle due to Niels Bohr: the correspondence principle. Essentially, according to this principle, every new physical theory must contain as a limit case the old theory. So, the Classical Mechanics must be a limiting case of Quantum Mechanics. How is that? Similitude again! The geometrical optics are obtained from the wave optics as a limiting case when the wavelength $\lambda \to 0$. It means that the phase $\varphi$ in (54) is a very big number and as a result $\psi(x,y,z) = ae^{i\varphi}$ is a slowly varying function. Similarly, we can consider a wave-function describing a physical system as being $\psi(x,y,z) = \Phi e^{i\varphi_q}$. How can we choose "the quantum phase" $\varphi_q$ so that at the classical limit $\varphi_q$ will be a big number? There is a constant existing all throughout Quantum Mechanics: $\hbar$. It is a very small quantity, if we look at it from the classical point of view, because it is of the order of $10^{-34}$. Choosing $\varphi_q$ to be $S/\hbar$, then the classical limit of Quantum Mechanics will be $\hbar \to 0$. Now, the wave-function for a quasi-classical physical system can be written as:

$$\Psi = \Phi e^{\frac{i}{\hbar}S} \tag{118}$$

Let us suppose that equation (118) is a solution of the Schrödinger equation (100). Introducing (118) in (100) and separating the real and the imaginary parts, we will have:

$$\left[\frac{(\nabla S)^2}{2m_0} + U + \frac{\partial S}{\partial t}\right]\Phi - \frac{\hbar^2}{2m_0}\Delta\Phi =$$
$$i\hbar\left[\frac{1}{m_0}(\nabla S \cdot \nabla \Phi) + \frac{\Phi}{2m_0}\Delta S + \frac{\partial \Phi}{\partial t}\right] \tag{119}$$

The classical limit means $\hbar \to 0$, hence the right part of (119) is null, which involves:

$$\frac{(\nabla S)^2}{2m_0} + U + \frac{\partial S}{\partial t} = 0 \tag{120}$$

What about S? From (119) we can see that S has the dimensions of *action,* as we know from Classical Mechanics [1,6,9], so, if S is the *action* of the physical system, then equation (120) is nothing else but the Hamilton − Jacobi equation of a particle of mass "m" moving in a potential U. Indeed, the Classical Mechanics is the limiting case of Quantum Mechanics, when $\hbar \to 0$. Also, Paul Ehrenfest [10] proved two theorems asserting: *the expected values of Quantum Mechanics' observables satisfy the same equations as Classical Mechanics' variables.* To prove Ehrenfest's theorems is a pretty easy task if we use the results of "*the expected values* and *the time derivative of the expected values*" discussed above.

*Conclusion*

The five postulates presented above, may be considered as the general frame of the conventional Quantum Mechanics and they will help in applying Quantum Mechanics to simple physical systems.